\documentclass[aps,10pt, twocolumn, groupedaddress, superscriptaddress, prl, nofootinbib]{revtex4-2}
\usepackage{amsmath}
\usepackage{mathtools}
\usepackage{amssymb}
\usepackage{graphicx}
\usepackage{textcomp}
\usepackage{bm}
\usepackage{soul}
\usepackage{subcaption}
\usepackage{cleveref}
\usepackage{bbold}
\usepackage[export]{adjustbox}
\usepackage{cancel}

\makeatletter
\newcommand*\bigcdot{\mathpalette\bigcdot@{.5}}
\newcommand*\bigcdot@[2]{\mathbin{\vcenter{\hbox{\scalebox{#2}{$\m@th#1\bullet$}}}}}
\makeatother

\DeclareMathOperator{\Tr}{Tr}

\newcommand{\braket}[1]{\langle{#1}\rangle}
\newcommand{\btf}[2]{|{#1}\rangle\langle{#2}|}

\usepackage[usenames,dvipsnames]{color}

\usepackage[braket, qm]{qcircuit}

\begin{document}

\title{Fully selective charging of a quantum battery by a purely quantum charger}

\author{Yohan Vianna}\email{yohan.vianna@lpmmc.cnrs.fr}
\affiliation{LPMMC, CNRS, Grenoble 38000, France}

\author{Marcelo F. Santos}\email{mfsantos@if.ufrj.br}
\affiliation{Instituto de Física, Universidade Federal do Rio de Janeiro, Rio de Janeiro 21941-972, Brasil}

\pacs{xxxx, xxxx, xxxx}                                         
\date{\today}
\begin{abstract}
In this paper we discuss a protocol for charging a two-level quantum battery using a bipartite charger composed of two quantum harmonic oscillators. As one of its features, it allows us to fully charge the battery and is universally optimal in the regime of a single excitation added as energy input. We also make use of a selective interaction to extend the protocol for a different class of quantum states and show that, in this case, the presence of quantum coherence can be harnessed as energetic resource to charge multiple similar batteries. Among these, we also explore symmetries of the derived effective dynamics to quickly discuss how the same protocol can be adapted to the task of \textit{active state resetting}, a task which is particularly useful in the quantum computation area.
\end{abstract}

\maketitle

\section{Introduction}

Quantum batteries are among the most basic, yet practical, pieces of technology present in the new century's toolbox of quantum gadgets. Despite its practical role in developing stable, \textit{on demand} sources of energy, the study of quantum batteries lies within a more fundamental realm of modern quantum mechanics: quantum energetics. In this scenario, a plethora of different investigations have been carried, either on the boosting of the battery's charging performance~\cite{Binder2015, Ferraro2018, sergi2020, Tacchino2020, FBarra2022, Vianna2023}, including the use of structured reservoirs~\cite{Lu2025, Tacchino2020B, Santos2021}, on the stability of the energy storage~\cite{cres2020, Santos2019, BM2020}, on the effects of real quantum resources~\cite{karen2013, Andolina2019, Kamin2020} present in the local or global system's state and, recently, on its applications as energy source for quantum computation~\cite{Kurman2026}.

The most prototypical model for a quantum battery comprises a (sometimes effective) two level system interacting with two or more quantum and/or classical systems. However, the broad range of physical implementations for quantum batteries span atomic systems in cavity QED~\cite{Buy2022, Horne2020, Klatzow2019}, superconducting systems~\cite{Chang2022, Elg2025, Hu2026}, and spin systems~\cite{Le2018, liu2021, Chand2025}, to name a few. In parallel to the interesting phenomena arising from these practical applications, quantum batteries also attract the attention from researchers of the quantum thermodynamics area~\cite{Barra2022, Garcia2020, Campaioli2024}, since these systems provide an intuitive playground for studying the energetic relations that are present in the interactions between two or more quantum systems.

The simplest way to charge a two-level quantum battery is by using an external classical drive as energy input. However, this leads to an unitary evolution of the battery's state which constrains its Von-Neumann's entropy and therefore limits the energy absorption from the charger. For example, if the qubit is initially in a thermal state, the best that can be done in this scenario is to swap the population between the ground and the excited states. To surpass this limitation an entropy dispenser is required. For instance, standard optical pumping represents a protocol that employs quantized external degrees of freedom to reduce the Von-Neumann's entropy of the qubit. In~\cite{Santos2023}, the authors use an auxiliary single mode oscillator as an entropy dispenser to beat both dynamics (unitary evolution and standard optical pumping) and increase the charging capacity of the battery.

In this work we introduce a charging scheme that employs two quantum harmonic oscillators that cooperate in order to charge a two-level quantum battery. This cross-mode charger leverages from an effective three-body interaction that allows for optimal charging with a single excitation as energetic resource. As it turns out, this effective dynamics can also be inverted and employed for resetting the state of the qubit to ground with high probability, a necessary task for quantum computation and one of DiVincenzo's criteria~\cite{Vincenzo2000} for a real quantum computer.

One of the problems of avoiding classical energy sources is that the correlations created between quantum batteries and quantum chargers in the course of the charging task typically limits the charging capacity of the former. Here, we tackle this issue using a selective interaction to extract the maximal amount of energy from the quantum charger.

This paper is divided as follows: in the first section we introduce the analysed model's full Hamiltonian and derive the effective dynamics in the dispersive regime. At this point we discuss the optimal regime for operation and point out its main limitation for the proposed task, which is naturally overcome in the following section where we expand the Hamiltonian by introducing the selectivity tool. This new addition allows us to consider a broader range of input energetic profiles and perform the protocol for multiple quantum batteries while extracting the maximal amount of energy used as input.

\section{Effective dynamics derivation}

\begin{figure}[h!]
\centering
\includegraphics[width=\columnwidth]{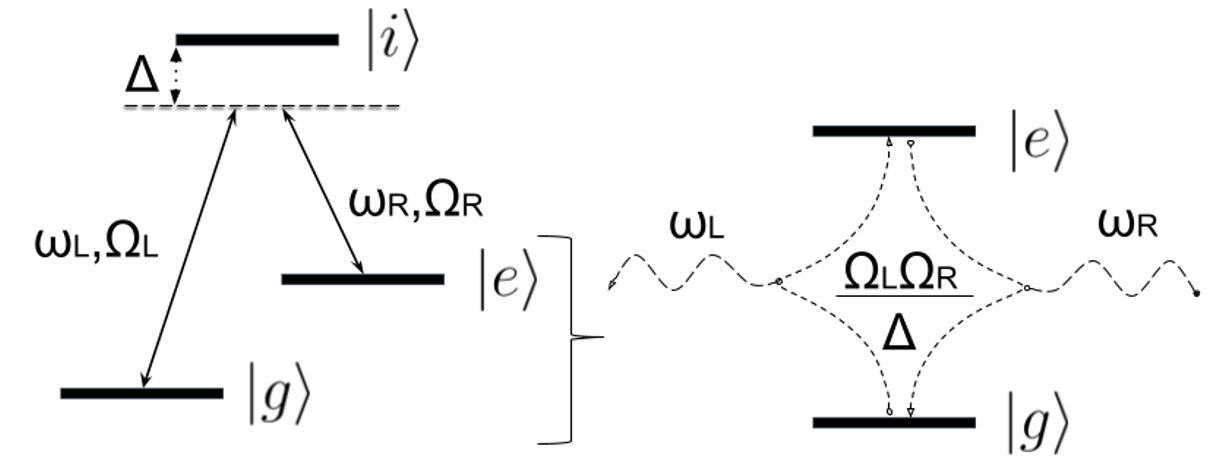}
\caption{Full dynamics (left) connecting two quantum harmonic oscillators to a three level system. The dispersive regime gives rise to an effective interaction mixing the three different subsystems (right).}
\label{scheme}
\end{figure}

The model employs a three level system (qutrit) in $\Lambda$ configuration, with free Hamiltonian $H_{0}^{\Lambda} = \sum_{j = g,e,i} \hbar\omega_{j}\hat{\sigma}_{jj}$, where $\omega_{g} < \omega_{e} < \omega_{i}$ and $\hat{\sigma}_{jk} = \btf{j}{k}$, interacting with two quantum harmonic oscillators of frequencies $\omega_{L} = \omega_{ig} - \Delta$ and $\omega_{R} = \omega_{ie} - \Delta$ ($\omega_{jk} = \omega_{j}-\omega_{k}$). Each oscillator is off-resonantly coupled to a different transition of the qutrit, both with an energy detuning $\hbar\Delta$, and independent coupling constants $\Omega_{L}$ and $\Omega_{R}$, as shown in Fig. (\ref{scheme}).

The unitary evolution of this system presents Rabi oscillations in both subspaces $\{\ket{g},\ket{i}\}$ and $\{\ket{e},\ket{i}\}$, which evolves an initial global state as $\rho_{0} \rightarrow \rho(t) = U(t)\rho_{0}U^{\dagger}(t)$, with $U(t) = \exp\left(-i\frac{t}{\hbar}H_{T}\right)$ and $H_{T}$ the system's total Hamiltonian given by

\begin{equation}
H_{T} = H_{0}^{\Lambda} + H_{0}^{L} + H_{0}^{R} + H_{int}^{\Lambda -L} + H_{int}^{\Lambda -R} + H_{drive}^{(M,N)}.
\end{equation}

Here $H_{0}^{k} = \hbar\omega_{k}\hat{a}^{\dagger}_{k}\hat{a}_{k}$ ($k = R, L$) is the free Hamiltonian of the $k-th$ oscillator and $H_{int}^{\Lambda -k}$ its interaction with the $\Lambda$ qutrit. For clarity, we write $H_{int}^{\Lambda -L} = \hbar\Omega_{L}\left(\hat{\sigma}_{gi}\hat{a}^{\dagger}_{L} + \hat{\sigma}_{ig}\hat{a}_{L}\right)$ and $H_{int}^{\Lambda -R} = \hbar\Omega_{R}\left(\hat{\sigma}_{ei}\hat{a}^{\dagger}_{R} + \hat{\sigma}_{ie}\hat{a}_{R}\right)$ in the standard rotating wave approximation for the dipole coupling between each oscillator and its respective transition. Furthermore, $H_{drive}^{(M,N)}$ represents an external classical drive added to the $\ket{g}\leftrightarrow\ket{e}$ transition, performing a two-parameter d.c Stark shift of the form:

\begin{equation}
H_{drive}^{(M,N)} = \left(\hbar\dfrac{\Omega_{R}^{2}}{\Delta}(N+1) - \hbar\dfrac{\Omega_{L}^{2}}{\Delta}M\right)\hat{\sigma}_{z},
\end{equation}

\noindent where $\hat{\sigma}_{z} = \hat{\sigma}_{ee}-\hat{\sigma}_{gg}$. This drive corrects the qubit's transition frequency $\omega_{eg}$ by an amount that depends on the couplings and on a pair of arbitrary but fixed integers $(M,N)$. Its purpose will be discussed later in this paper.

For the sake of this work, we are mainly interested in the dispersive regime $\Delta \gg \Omega_{L},\Omega_{R}$, when rapid oscillating degrees of freedom average out to an effective dynamics separated for level $|i\rangle$ and the $\{\ket{g} ,\ket{e}\}$ subspace~\cite{Etienne2007} and dictated by the interaction picture Hamiltonian $\tilde{H}_{eff} = \hbar\left[\dfrac{\Omega_{L}^{2}}{\Delta}\left(\hat{a}^{\dagger}_{L}\hat{a}_{L} + 1\right) + \dfrac{\Omega_{R}^{2}}{\Delta}\left(\hat{a}^{\dagger}_{R}\hat{a}_{R} + 1\right)\right]\hat{\sigma}_{ii} + H_{LqR}$. Here, $H_{LqR}$ provides the effective coupling between levels $\ket{g}$ and $\ket{e}$ and is given by

\begin{equation}
\begin{split}
H_{LqR} = &-\hbar\dfrac{\Omega_{L}^{2}}{\Delta}\left(\hat{a}^{\dagger}_{L}\hat{a}_{L}-M\right)\hat{\sigma}_{gg}\\
&- \hbar\dfrac{\Omega_{R}^{2}}{\Delta}\left(\hat{a}^{\dagger}_{R}\hat{a}_{R}-[N+1]\right)\hat{\sigma}_{ee}\\
&-\hbar\dfrac{\Omega_{L}\Omega_{R}}{\Delta}\left(\hat{a}_{L}\hat{\sigma}_{eg}\hat{a}^{\dagger}_{R} + \hat{a}^{\dagger}_{L}\hat{\sigma}_{ge}\hat{a}_{R}\right).
\end{split}
\label{QBHamiltonian}
\end{equation}

As we notice in $\tilde{H}_{eff}$, the level $\ket{i}$ is adiabatically eliminated from the dynamics and only serves as a virtual connection between levels $\ket{g}$ and $\ket{e}$. Moreover, this effective dynamics takes place within a time scale dictated by the coupling $\lambda_{eff} = \frac{\Omega_{L}\Omega_{R}}{\Delta}$, in which one quantum of energy is transferred between the two oscillators, altering the state of the effective qubit in the process.

The Hamiltonian $H_{LqR}$ has two trivial sets of eigenstates $\ket{0,g,n}$ and $\ket{m,e,0}\ (m,n\in\mathbb{N})$, where we use the notation $\ket{m,j,n} \equiv \ket{m}_{L}\otimes\ket{j}\otimes\ket{n}_{R}$ to indicate the global state with $m$ excitations on the left oscillator, the qubit in state $\ket{j}$ ($j=g,e$) and $n$ excitations on the right oscillator. Moreover, $H_{LqR}$ is block diagonal in the subspaces $\{\ket{m,g,n},\ket{m-1,e,n+1}\}$ $(m\geq 1, n\geq 0)$, thus making each doublet evolve in time as $\ket{m,g,n} \rightarrow A_{mn}(t)\ket{m,g,n} + B_{mn}(t)\ket{m-1,e,n+1}$ and $\ket{m-1,e,n+1} \rightarrow B_{mn}(t)\ket{m,g,n} + A^{\ast}_{mn}(t)\ket{m-1,e,n+1}$, where we define the geometric functions of the evolution $A_{mn}(t) = \cos\left(\frac{\Omega_{mn}t}{2}\right) + i\frac{\Delta_{mn}}{\Omega_{mn}}\sin\left(\frac{\Omega_{mn}t}{2}\right)$ and $B_{mn}(t) = i \frac{\sqrt{\Omega_{mn}^{2} - \Delta_{mn}^{2}}}{\Omega_{mn}}\sin\left(\frac{\Omega_{mn}t}{2}\right)$, with $\Delta_{mn} = \lambda_{eff}\left(\chi^{-1}(m-M) - \chi (n-N)\right)$ the doublet detuning, $\Omega_{mn} = \sqrt{4m(n+1)\lambda_{eff}^{2} + \Delta_{mn}^{2}}$ its Rabi frequency, and $\chi \equiv \Omega_{R}/\Omega_{L}$ the ratio between the oscillators' couplings.

Some interesting physical features of this Hamiltonian arise from the aforementioned sets of trivial eigenstates: the set $\ket{m,e,0}\ \forall m$ ($\ket{0,g,n}\ \forall n$) is responsible for keeping a share of the population of the qubit's excited (ground) state shielded against the time evolution. This shielding is manifestly dependant on the amount of vacuum population present in the right (left) oscillator, regardless of the state of the left (right) one. In the next section we propose two useful tasks and analyse the general dynamics to find the optimal energy input profile for each one. For clarity, we initially consider the simpler regime where $\chi = 1$, \textit{i.e.}, the individual couplings satisfy $\Omega_R = \Omega_L \ll \Delta$, and remove the external drive by setting $M=0$, $N=-1$. Later on we extend the analysis to consider a broader range of values for $\chi$ and discuss how $H^{(M,N)}_{drive}$ is employed in this extension.\\

\section{Protocol for charging or resetting}

As previously mentioned, two useful tasks involving a two-level system are charging and resetting~\cite{Kobayashi2023, Diniz2023, DeCross2023} its state. The charging task is a process where the qubit, thought of as a quantum battery, interacts with another system named \textit{charger} (in our case, the bimodal harmonic oscillators) and evolves within a certain time window $[0,\tau]$ such that its final state $\rho^{q}_{\tau}$ is more energetic than the initial one $\rho^{q}_{0}$. The difference in internal energy at the end of the process, $\Delta \mathcal{U}^{q}(\tau)$, is maximal for the optimal regime of operation of the protocol. Conversely, resetting the qubit's state can be thought as fully discharging it by transferring energy from the qubit to the charger. In this section we analyse the effective dynamics presented in the previous section to find the optimal initial state of the charger that performs each of these tasks.

We start by considering a global initial state of the form $\rho (0) = \rho^{L}\otimes\rho^{q}\otimes\rho^{R}$, \textit{i.e.}, with all subsystems uncorrelated to each other. We also consider that the initial state of the qutrit commutes with its free Hamiltonian, therefore eliminating any possible coherent resource initially present in it. Under these general assumptions plus the adiabatic elimination of level $|i\rangle$, the variation of the internal energy of the qubit (subsystem $\{\ket{g} ,\ket{e}\}$) is given by

\begin{equation}
\dfrac{\Delta \mathcal{U}^{q}(\tau)}{2\hbar\omega_{eg}} = \sum_{\substack{m=1 \\ n=0}}^{\infty}\left(p^{L}_{m}p^{q}_{g}p^{R}_{n} - p^{L}_{m-1}p^{q}_{e}p^{R}_{n+1}\right)|B_{mn}(\tau)|^{2},
\label{deltaUq}
\end{equation}

\noindent where $p^{I}_{j} \equiv \bra{j}\rho^{I}\ket{j}$ are the initial populations for each subsystem and $B_{mn}(t)$ is the geometric function defined in the previous section. 

In order to find out the optimal initial state for the bimodal charger for each task, let us consider the raw energy matrix elements $S_{mn} \equiv p^{L}_{m}p^{q}_{g}p^{R}_{n} - p^{L}_{m-1}p^{q}_{e}p^{R}_{n+1}$, that are proportional to the maximum amount of energy that can be exchanged in each doublet during the time evolution. Since the geometric factor $|B_{mn}(\tau)|^{2}$ in the energy variation of the qubit is always non-negative, the optimal initial state of the charger for increasing (reducing) the energy of the qubit satisfies $S_{mn} > 0$ ($S_{mn} < 0$) for each and every pair $\{m\geq 1, n\geq 0\}$. By summing over $\{m,n\}$ in this domain, we find that our protocol is optimal in charging (resetting) the qubit's state whenever the quantity

\begin{equation}
\sum_{\substack{m=1 \\ n=0}}^{\infty} S_{mn} = p^{q}_{g}\left(1 - p^{L}_{0}\right) - p^{q}_{e}\left(1 - p^{R}_{0}\right)
\label{rawenergy}
\end{equation}

\noindent is maximally positive (negative). Therefore, the optimal initial state for charging the qubit is the one where $p^{L}_{0} = 0$ and $p^{R}_{0} = 1$. Conversely, qubit reset in our protocol is optimal when $p^{L}_{0} = 1$ and $p^{R}_{0} = 0$. This can be understood in the light of the aforementioned features of the trivial set of eigenstates: as the vacuum population of the right oscillator increases, so does the amount of excited population of the qubit which is locked throughout the dynamics; therefore, the optimal regime in our protocol for charging the qubit is met whenever the right oscillator is purely vacuum. The same line of reasoning can be carried to the task of qubit resetting regarding the vacuum population on the left mode.

\subsection{Universal optimization}

Given any arbitrary state of a qubit, its initial internal energy can be written as $\mathcal{U}^{q}_{0} = \hbar\omega_{eg}\left(p^{q}_{e} - p^{q}_{g}\right)$. Any universally optimal protocol for charging such qubit should drive its internal energy to its maximum, namely $\mathcal{U}^{q}_{\tau} = \hbar\omega_{eg}$, which means an energy variation of $\Delta \mathcal{U}^{q}(\tau) = 2p^{q}_{g}\hbar\omega_{eg}$.

As it turns out from the previous analysis, the optimal initial states for both modes in the charger must be of the form $\rho^{L} = \sum_{m=1}^{\infty}p^{L}_{m}\btf{m}{m}$ and $\rho^{R} = \btf{0}{0}$. If we choose to perform the charging task with minimal energy input that means choosing for the $L$ oscillator a population distribution where only a single excitation is present, with the minimal energy $\hbar\omega_{L}$. If we input $p^{L}_{n} = \delta_{n1}$ and $p^{R}_{m} = \delta_{m0}$ in Eq. (\ref{deltaUq}), then on time $t = \tau = \pi/\Omega_{10}$ we find precisely $\Delta \mathcal{U}^{q} (\tau) = 2p^{q}_{g}\hbar\omega_{eg}$, which establishes our protocol as universally optimal in the regime of minimal energy input. Naturally, a similar reasoning can be carried for the task of qubit resetting, leading to the same conclusion.

\subsection{Resource consumption under realistic scenarios}

From the results derived so far we have seen that purifying one oscillator to its vacuum state and preparing a pure single excitation on the other is the condition for optimal charging or resetting of the qubit. Strictly speaking, however, that can only happen at zero temperature. In this section we discuss the application of our protocol in a more realistic scenario, where temperatures are taken into account.

\begin{figure}[t]
	\centering
    \includegraphics[width=\linewidth]{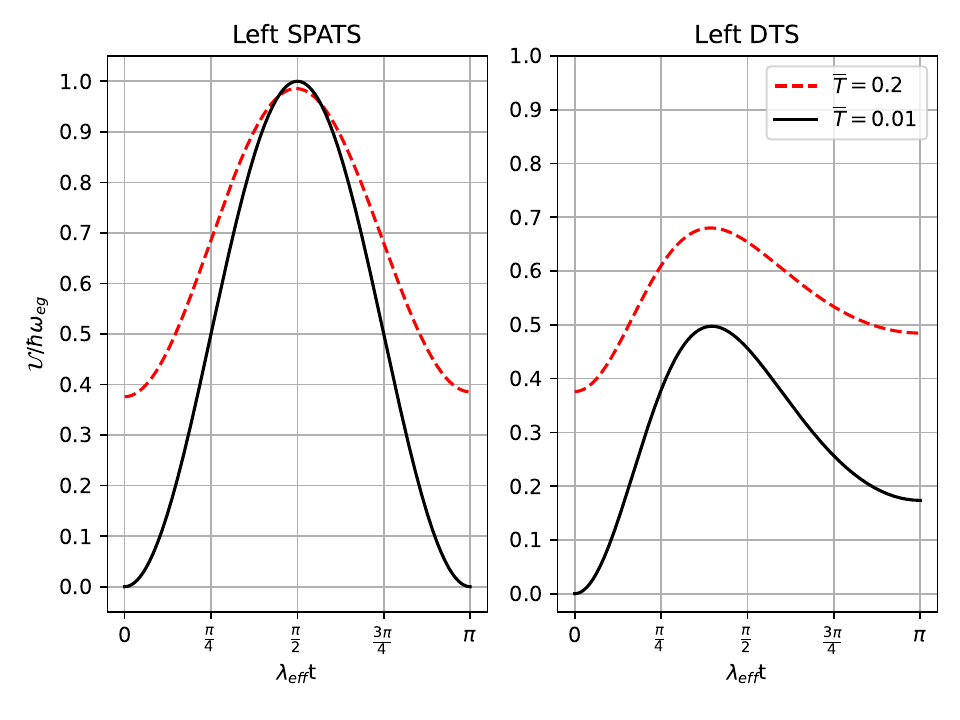}
    \caption{The temperature range for the optimal operation of our protocol is mainly limited by the presence of initial population in the level $\ket{i}$ of the qutrit, which does not evolve in the time scale of the effective evolution and therefore does not contribute to the battery's energy gain. In these plots we present the time evolution of the internal energy in the effective qubit for the two extremal temperatures in the allowed range. As it can be seen, the SPATS scenario makes the protocol robust against temperature variations in this range, in comparison to the DTS case.}
    \label{Erg}
\end{figure}

From the point of view of resource consumption, the usual goal is to optimally perform some task with the least amount of resources possible. In this scenario, thermal states are considered as ``free" resources, in the sense that thermalization of a quantum state occurs naturally, with no need of extra intervention by an experimentalist. Therefore, a more realistic consideration for our protocol would be that all three subsystems are initially in thermal equilibrium with some finite temperature reservoir, \textit{i.e.}, $\rho^{I} = \rho^{I}_{\overline{T}}$ is the Gibbs state for system $I$ with the dimensionless temperature of the reservoir $\overline{T} = k_{B}T/\hbar\omega_{ig}$ and average number of excitations $\langle n\rangle_{\overline{T}} = \Tr\left(\hat{a}^{\dagger}_{I}\hat{a}_{I}\rho^{I}_{\overline{T}}\right)$. Notice that for this choice of initial state the raw energy matrix $S_{mn}$ is identically zero for all active doublets, and therefore the global system does not evolve unless some change is actively made. From now on we consider the charging protocol and leave as a side note that everything discussed can be trivially carried to the resetting task.

From the considerations discussed in the previous section, we seek a modification to the initial thermal states which maximizes the vacuum population on the right oscillator and minimizes it on the left one, but with minimal energy addition to the latter. Even though perfect vacuum is only achieved in the ideal limit of zero temperature, the fact that the right oscillator couples to the $\ket{e}\leftrightarrow\ket{i}$ transition ($\omega_R \sim \omega_{ei}$) plays in our favor. As long as $\omega_{eg}/\omega_{ie}\ll 1$ and the temperature is not too high, the initial ground state population of the qubit $p^g_T$ is much smaller than the initial vacuum population of the right oscillator, i.e. for all practical purposes, as long as the temperature is not too high, the right oscillator will be very close to its vacuum state when compared to the qubit, and one should expect almost perfect locking of the population of level $\ket{e}$ in the limit $\omega_{eg}/\omega_{ie}\ll 1$. Indeed, this limit is also necessary to guarantee no spurious populations in level $\ket{i}$ in order to properly isolate the qubit subspace from it.

On the left oscillator mode, however, the presence of the vacuum is undesirable for the charging task since it would also lock a high portion of the ground state's population. For the reasons discussed in the previous section, the best modification to be made is then to coherently add a single excitation in the previously prepared Gibbs state, thus generating a single-photon added thermal state (SPATS). This state is defined for the left mode as $\rho^{L}_{SPATS} = \hat{a}_{L}^{\dagger}e^{-\overline{\beta} H_{0}^{L}}\hat{a}_{L}/Z_{1}$, with $\overline{\beta} = 1/\overline{T}$ the inverse dimensionless temperature defining the thermal reservoir and $Z_{1}$ the partition function for the SPATS. The average number of excitations present in this state is $\braket{n}_{SPATS} = 2Z_{1} - 1$, which is commensurate with the Gibbs state average number of excitations (check Appendix). This means that, for the regime of temperatures we are interested in this paper, the amount of energy added to the initial Gibbs state is approximately one unit of $\hbar\omega_{L}$.

Adding a single photon may present some experimental challenges depending on the specific setup. Therefore, we have also considered a simpler way to add energy to the left oscillator that is to couple it to a coherent source in order to produce a displaced thermal state (DTS) in it. In particular, we investigate how much charge a DTS would provide to the effective qubit if restricted to the same amount of input energy of the SPATS. The DTS is defined for the left mode as $\rho^{L}_{DTS} = \hat{D}(\alpha)\rho^{L}_{\overline{T}}\hat{D}(\alpha)^{\dagger}$, where $\hat{D}(\alpha)$ is the displacement operator and $\alpha$ satisfies $|\alpha |^{2} + \langle n\rangle_{\overline{T}} = \langle n\rangle_{SPATS}$ (\textit{i.e.}, we keep $\alpha$ low enough so that the average number of excitations in the DTS is the same as in the SPATS case). 

In Fig. \ref{Erg} we present the time evolution for the effective qubit's internal energy for the two different scenarios under two different temperatures. Note that the SPATS performs much better than the DTS for the same added energy to the charger. This is easy to explain: when restricted to a low energy injection, i.e. a small displacement $\alpha$, the vacuum population of the DTS is still too high and more ground state of the qubit gets locked by the interaction.

\begin{figure}[t]
\centering
\includegraphics[width=\linewidth]{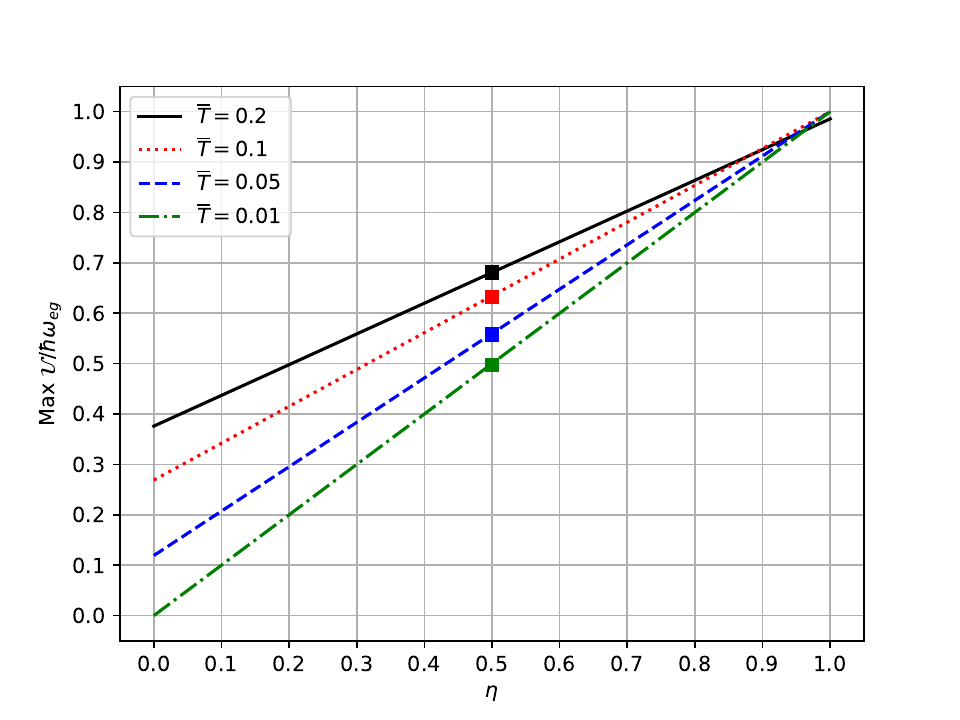}
\caption{Qubit's maximal internal energy for an initial inefficient SPATS left mode, as a function of its efficiency of creation. The dots represent the same quantity for the DTS initial state at the respective temperature.}
\label{FuncEta}
\end{figure}

However, as previously mentioned, in many experimental setups, displacing a thermal state is much simpler than producing a SPATS. In fact, in the most common quantum optical protocols, the displacement process involves deterministically combining a travelling pseudo-thermal field with an attenuated laser in a beam splitter\cite{Paris1996, Guo2018, Walton2024}, whereas the coherent photon addition usually relies on probabilistic non-linear processes and heralding measurements\cite{Biagi2021, Fadrny2024, Kiesel2011, Zavatta2017}. Defining $\eta$ as the net efficiency for successfully adding a single excitation to a thermal oscillator, we model the initial state of the left mode as the convex sum $\rho^{L}(\eta) = \eta\rho^{L}_{SPATS} + (1-\eta)\rho^{L}_{\overline{T}}$. Figure \ref{FuncEta} displays the maximal energy stored in the effective qubit as a function of $\eta$ for different temperatures. Notice that, for the whole range of possible temperatures, the inefficient SPATS initial state surpasses the DTS at the respective temperature precisely at $\eta = 50\%$. Deviations from this behaviour are expected for higher temperatures, but in this case associated to non-negligible population trapping on level $\ket{i}$.

\subsection{Open dynamics}

So far, we have carried the analysis solely at the unitary level, \textit{i.e.}, assuming that the coupling to the thermal environment is small enough to produce negligible dynamics at the desired time scale. In realistic scenarios, however, that is rarely the case and the presence of the thermal environment should be accounted for as it spoils coherent dynamics in the system by destroying quantum correlations between subsystems that are paramount for the described protocols. In this section we analyse the overall evolution of the system when it interacts dissipatively with a Markovian thermal reservoir at temperature $\overline{T}$. The global evolution in this scenario is modelled by a Lindblad master equation of the type

\begin{equation}
\dot{\rho}(t) = -i\left[H_T,\rho(t)\right] + \sum_{\mu}\gamma_{\mu}^{+}\mathcal{D}_{\mu}^{+}\left[\rho (t)\right] + \sum_{\mu}\gamma_{\mu}^{-}\mathcal{D}_{\mu}^{-}\left[\rho (t)\right]
\label{MarkovianLindblad}
\end{equation}

\noindent where the summations are carried over the indices $\{\mu\} = \{ig,\ ie,\ L,\ R\}$ and the dissipators $\mathcal{D}_{\mu}^{\pm}\left[\bigcdot\right] = \hat{l}^{\pm}_{\mu}\bigcdot\hat{l}_{\mu}^{\pm\dagger} - \frac{1}{2}\{\hat{l}_{\mu}^{\pm\dagger}\hat{l}^{\pm}_{\mu},\bigcdot\}$. Here, $\hat{l}^{+}_{\mu}$ ($\hat{l}^{-}_{\mu}$) is the raising (lowering) operator of one excitation in the transition $\mu$ as in $\{\hat{l}^{+}_{\mu}\} = \{\hat{\sigma}_{ig},\hat{\sigma}_{ie},\hat{a}^{\dagger}_{L},\hat{a}^{\dagger}_{R}\}$ and $\{\hat{l}^{-}_{\mu}\} = \{\hat{\sigma}_{gi},\hat{\sigma}_{ei},\hat{a}_{L},\hat{a}_{R}\}$. Moreover, the dissipation rates $\gamma_{\mu}^{\pm}$ are those for a thermal environment, \textit{i.e.}, $\gamma_{\mu}^{+} = \gamma_{0\mu}\overline{n}(\omega_{\mu})$ and $\gamma_{\mu}^{-} = \gamma_{0\mu}\left(\overline{n}(\omega_{\mu})+1\right)$, where $\overline{n}(\omega_{\mu}) = (e^{\hbar\omega_{\mu}/k_{B}T}-1)^{-1}$ is the Bose-Einstein distribution of excitations with mean energy $\hbar\omega_{\mu}$ at the environmental temperature $T$. For simplicity we fix $\gamma_{0\mu} = \gamma_{0}$ constant for all the involved transitions, and parametrize it in terms of the effective coupling $\lambda_{eff}$. Finally, we will consider that, due to selection rules, the dissipation occurring in the transition $\ket{e}\leftrightarrow\ket{g}$ is weaker than the others and takes place in a different time scale, and therefore we omit it.

\begin{figure}[t]
  \centering
  \begin{subfigure}[b]{0.48\textwidth}
    \centering
    \includegraphics[width=\linewidth]{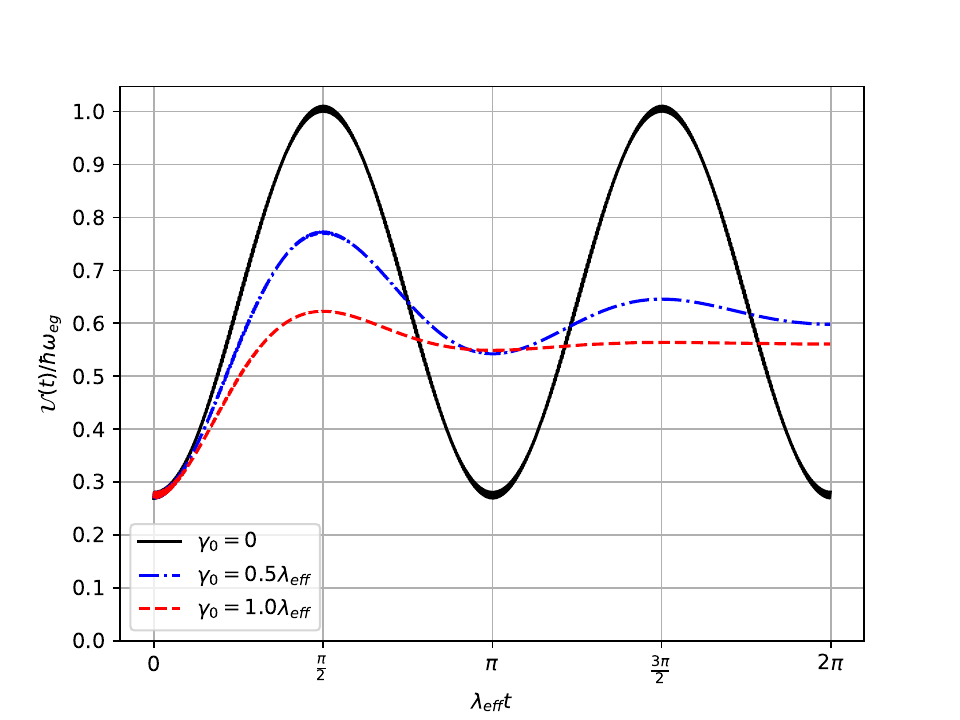}
  \end{subfigure}
  \vspace{0.5cm}
  \begin{subfigure}[b]{0.48\textwidth}
    \centering
    \includegraphics[width=\linewidth]{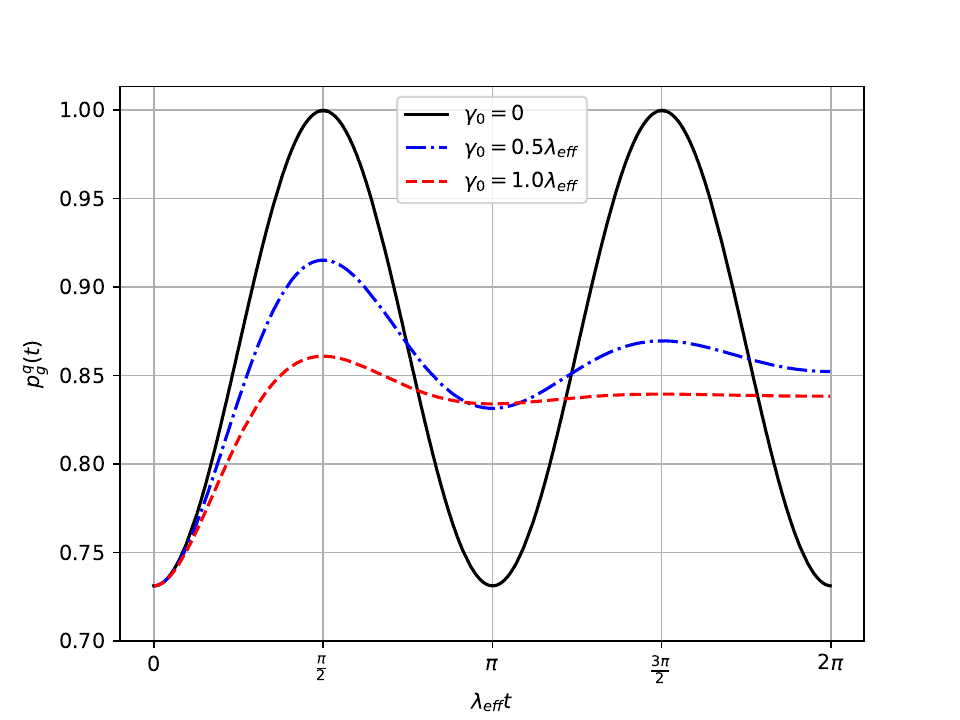}
  \end{subfigure}
  \caption{Qubit's internal energy in the charging protocol (upper) and ground level population in the resetting protocol (lower). Both plots are evaluated for different bare coupling strength's values and dimensionless temperature $\overline{T} = 0.1$.}
  \label{dissipation}
\end{figure}

The presence of dissipation in the whole system has two competing effects: for the qutrit, it spoils the charging by destroying the quantum correlations between qutrit and oscillators, which are necessary for population inversion in the qubit subspace. For the oscillators, however, dissipation drags the average number of excitations down to the point where the interaction Hamiltonian is not relevant for the evolution of the global system and the effective qubit achieves a stable asymptotic state.

Figure \ref{dissipation} displays the evolution of the qutrit's internal energy for the scenario of best charge performance (\textit{i.e.}, when the single excitation added flows from the left to the right mode), as well as the ground level population (\textit{i.e.}, when the excitation flows from right to left). Notice that Markovian thermalization affects both protocols differently: in spite of the intrinsic $\{e\leftrightarrow g, R\leftrightarrow L\}$ symmetry present in the Hamiltonian, thermalization does not share the same property. A simple way to understand it is by recalling that the optimal scenario happens  when one of the oscillators is in vacuum and the other is away from it: in this sense, the entropy dispenser mode will always be favored by the presence of the Markovian reservoir, while the energy input mode will suffer from its deleterious effects. On the other hand, thermalization on the qubit always favors its ground state population more than the excited one, by dragging this subsystem down to its equilibrium state and draining all possible energy input in it. This simple reasoning shows why thermalization can be less deleterious for qubit resetting than for battery charging.

\section{Charging multiple batteries}

In the previous section we presented a straightforward protocol to charge (or reset) a qubit with minimal resource usage which, for optimal operation, relied on the equal couplings between the qutrit and the two oscillators composing the charger. However, the assumption of commensurate couplings is too restrictive for realistic applications, and therefore we seek to expand our protocol to encompass less idealistic coupling constants. For this matter, we make use of the selectivity tool which is physically implemented by the external d.c. drive~\cite{Santos2001,Santos05,Santos2023} and parametrically depends on a pair of arbitrary but fixed integers $(M,N)$. In what follows we quickly present how this mechanism works and how it can be used to extract maximal energy from any quantum state used as input in the charger.

\subsection{Selectivity regime}

Following the Hamiltonian (\ref{QBHamiltonian}), we now consider a broader range of values for the parameter $\chi = \Omega_{R}/\Omega_{L}$. In this scenario, the role of the controllable parameters $M, N$ is non-negligible and can be understood by looking at the non-trivial eigenvalues of $H_{LqR}$:

\begin{equation}
\dfrac{E_{\ket{m,g,n}}}{\hbar\lambda_{eff}} = \dfrac{M-m}{\chi},
\end{equation}

\begin{equation}
\dfrac{E_{\ket{m-1,e,n+1}}}{\hbar\lambda_{eff}} = \chi(N-n).
\end{equation}

This is enough to provide an understanding of the selectivity tool in our scheme: whenever $\chi \ll 1$ ($\chi \gg 1$) all doublets are highly out of resonance and therefore do not change considerably in the time scale given by $\lambda_{eff}^{-1}$ except the ones for which $m=M$ ($n=N$). Thus, with this tool, we not only make sure that the necessary resonance is triggered for the evolution to occur, but we also have the perk of controllability over which family of doublets are going to exchange excitations in the effective dynamics by exploiting any unavoidable asymmetry between the couplings. Also, notice that the external drive implementing the shift does not affect the conditions for the adiabatic elimination of level $\ket{i}$.

This extra degree of control provided by the selective operation allows us to surpass the vacuum barriers present in the initial state of the charger and extend the application of the effective interaction to other initial states. For the charging task, for instance, this barrier is naturally avoided by using a SPATS initial state on the left oscillator, but one may wonder whether it is possible to also extract as much energy from other types of states under this protocol's interaction. In what follows we present and discuss this paradigm in the context of charging a string of initially empty (thermal) qutrit batteries by employing a DTS initial state on the left mode.

\subsection{The multi-charge protocol}

As previously stated, we are now interested in exploring the selective interaction to benchmark the performance of the protocol in charging a string composed of K uncorrelated qutrits interacting one at a time with the charger. This collisional model assists us in understanding whether there is any intrinsic limitations related to the non-null vacuum population in the left oscillator for this particular task. To this extent, the charger's state at the beginning of each collision is given by

\begin{equation}
\rho^{RL} = \sum_{m,m'}\sum_{n,n'}\ p(m,m',n,n')\ket{m,n}\bra{m',n'},
\end{equation}

\noindent where $p(m,m',n,n') = \bra{m,n}\rho^{RL}\ket{m',n'}$ is the joint probability distribution for two quantum harmonic oscillators to transition from state $\ket{m,n}$ to $\ket{m',n'}$. Naturally, before the first collision both oscillators are separable and satisfy $p(m,m',n,n') = p^{L}(m,m')\times p^{R}(n,n')$, and we initialize the protocol with $p^{L}(m,m') = \bra{m}\rho^{L}_{DTS}\ket{m'}$ and $p^{R}(n,n') = \bra{n}\rho^{R}_{\overline{T}}\ket{n'}$. However, this separability condition is lost after the first battery gets maximally charged because the evolution correlates both subsystems. Moreover, we calibrate the flying time of the batteries such that the charging process is optimal for maximally charging each one of them.

\begin{figure}[h!]
  \centering
  \begin{subfigure}[b]{\columnwidth}
    \centering
    \includegraphics[width=\linewidth]{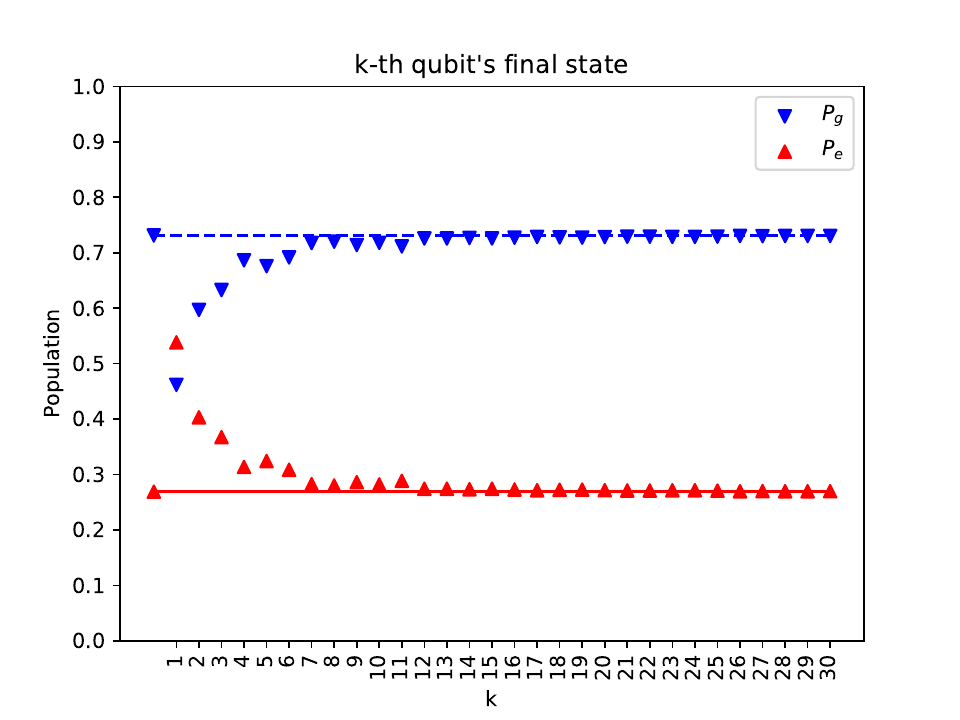}
    \label{collisions:a}
  \end{subfigure}
  \begin{subfigure}[b]{\columnwidth}
    \centering
    \includegraphics[width=\linewidth]{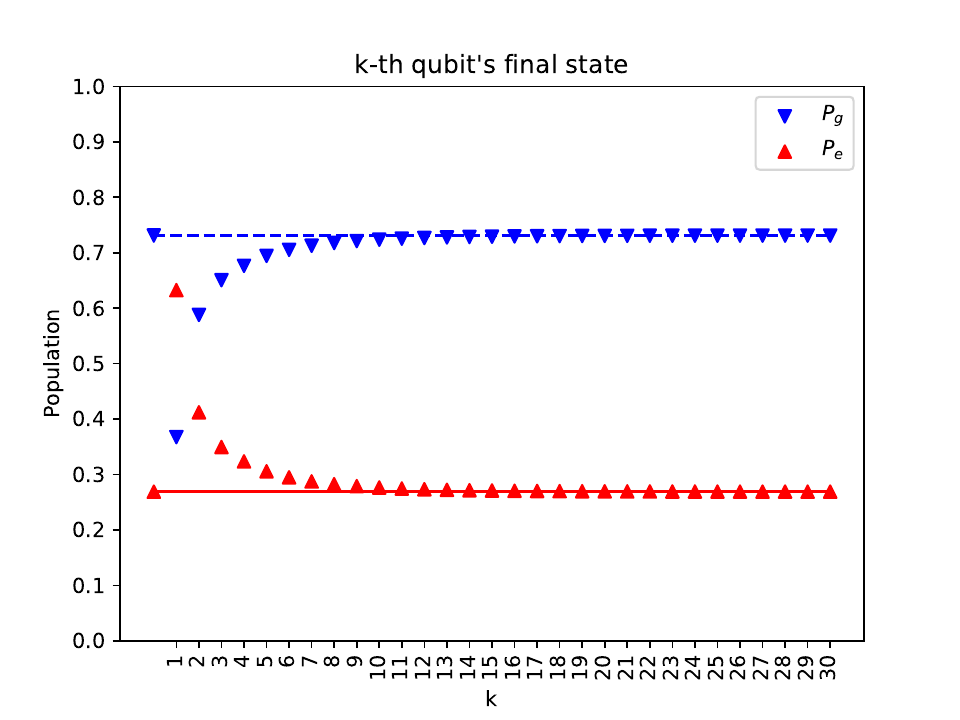}
    \label{collisions:b}
  \end{subfigure}
  \begin{subfigure}[b]{\columnwidth}
    \centering
    \includegraphics[width=\linewidth]{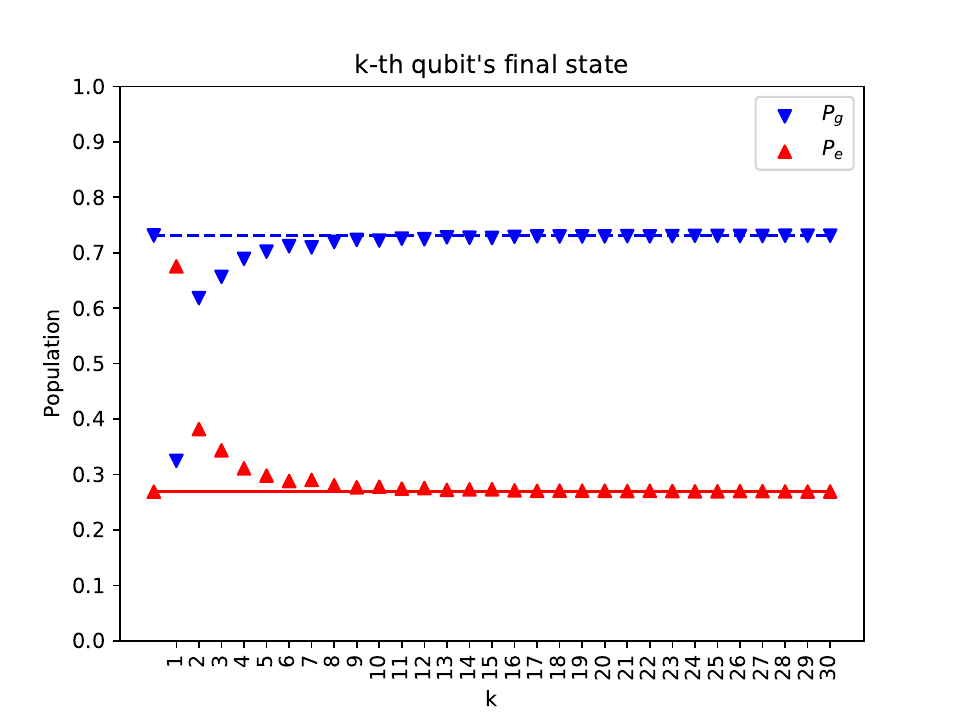}
    \label{collisions:c}
  \end{subfigure}
  \caption{Final state of each qubit after interacting with the charger. The three plots show the regimes $\chi\ll 1$, $\chi = 1$, and $\chi \gg 1$, respectively. The leftmost points, as well as the drawn lines, represent the thermal population of each qubit before interaction and are fixed by the dimensionless temperature $\overline{T} = 0.1$.}
  \label{collisions}
\end{figure}

Our goal is to extract the maximum amount of energy from a low-$\alpha$ displaced thermal state, for which the population of the vacuum is far from negligible, and for that we set the same small displacement as we did in the non-selective protocol (\textit{i.e.}, $\alpha$ being low enough to match the same amount of internal energy as the SPATS). The three possible domains for the coupling ratio $\chi$ are explored and in each scenario the parameters $(M,N)$ are chosen in each collision in order to select the family of doublets that maximizes the energy extraction by the colliding battery. The results for $K=30$ collisions are shown in figure \ref{collisions} for the final populations of each qubit after its collision and maximal energy extraction.

Notice that for all three scenarios population is only inverted for the first qubit. Also notice that the energy absorption in the following batteries modifies their population distribution towards Gibbs states with higher temperatures. This, for instance, represents energy that is accounted as resource in the context of resource theories~\cite{Gour2015, Brandao2013}. Another noteworthy point in this graph is that the energy acquired by each qubit is proportional to the height of its excited population after the collision in comparison to the initial (full, red) line. From this one can easily see that the total amount of energy extracted from the DTS in the multi-shot charging surpasses the energy extracted from the SPATS whenever it is possible to perform the selective operation with multiple batteries, thus establishing no fundamental limitations in using the SPATS in detriment of the DTS.

\section{Discussion}

At first sight, the asymptotic behaviour of the multiple charging protocol observed in figure \ref{collisions} seems to indicate the end of the full charging cycle, where the protocol is unable to further extract energy from the charger. A simple way to check this is by recalculating the total raw energy sum from all the active doublets.

\begin{figure}[h!]
\centering
\includegraphics[width=\linewidth]{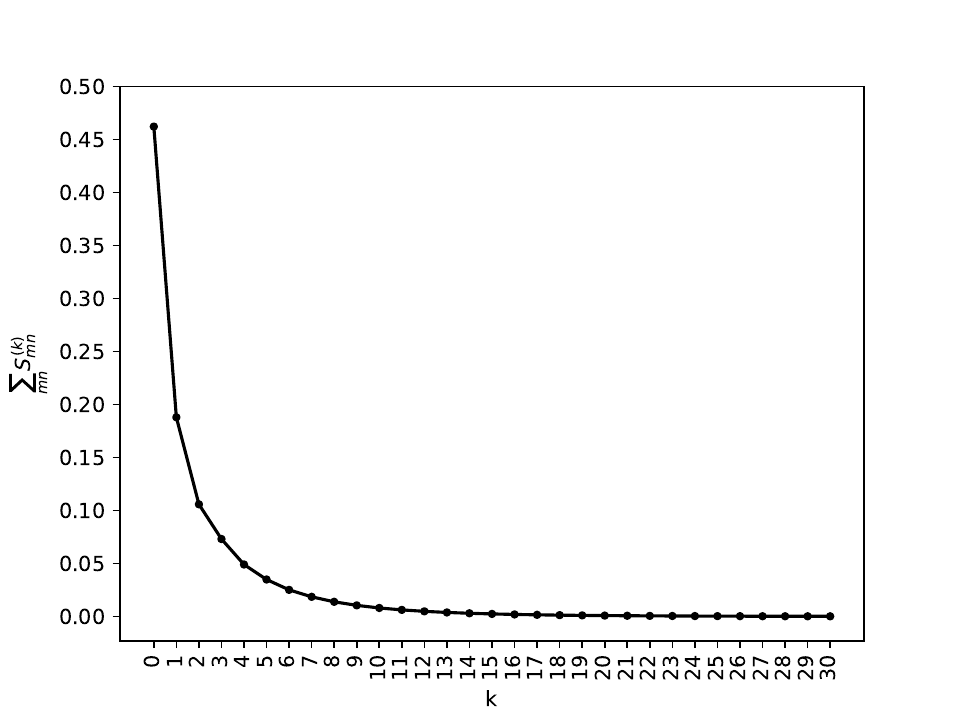}
\caption{Charger's total raw energy in each collision. The asymptotic behaviour shows that the multi-charging protocol indeed extracts the maximal amount of energy to the chain of batteries.}
\label{zeroenergy}
\end{figure}

For the generalized protocol, the raw energy matrix elements are defined as $S_{mn}^{(k)} =  p^{q}_{g}p_{m,m,n,n}^{(k,0)} - p^{q}_{e}p_{m-1,m-1,n+1,n+1}^{(k,0)}$ where we make use of the shorter notation $p_{m,m',n,n'} = p(m,m',n,n')$ for the joint fields' distribution and the index $(k,0)$ indicates the state of the charger at the beginning of the $k-th$ collision (which shall be suppressed from now on, for the sake of notation). Naturally, $p^{q}_{j}$ ($j=g,e$) are the initial populations of the qubit, which are the same for all colliding qubits in the chain and therefore do not carry the index $(k,0)$. Once more we call the attention for the fact that achieving a separable Gibbs state of the modes is a sufficient, yet not necessary, condition to make all elements of $S_{mn}^{(k)}$ vanish. In fact, as we soon prove, the asymptotic state of the charger is non-separable. Still, if we sum over all the active doublets elements of $S_{mn}^{(k)}$ we find that the total raw energy that can be extracted from the charger in each collision $k$ is given by $\sum_{m,n}S^{(k)}_{mn} = p^{q}_{g}\left(1-p^{L}_{00}\right)-p^{q}_{e}\left(1-p^{R}_{00}\right)$, indicating that even in the generalized protocol, where the modes in each collision do not start from a separable state, it is still the non-vacuum populations of each mode separately that controls the energetic dynamics. Figure \ref{zeroenergy} displays exactly the evolution of this quantity in each collision and shows that indeed in the asymptotic regime all the energy extractable from the charger has indeed been extracted.

\begin{figure}[h!]
\centering
\includegraphics[width=\linewidth]{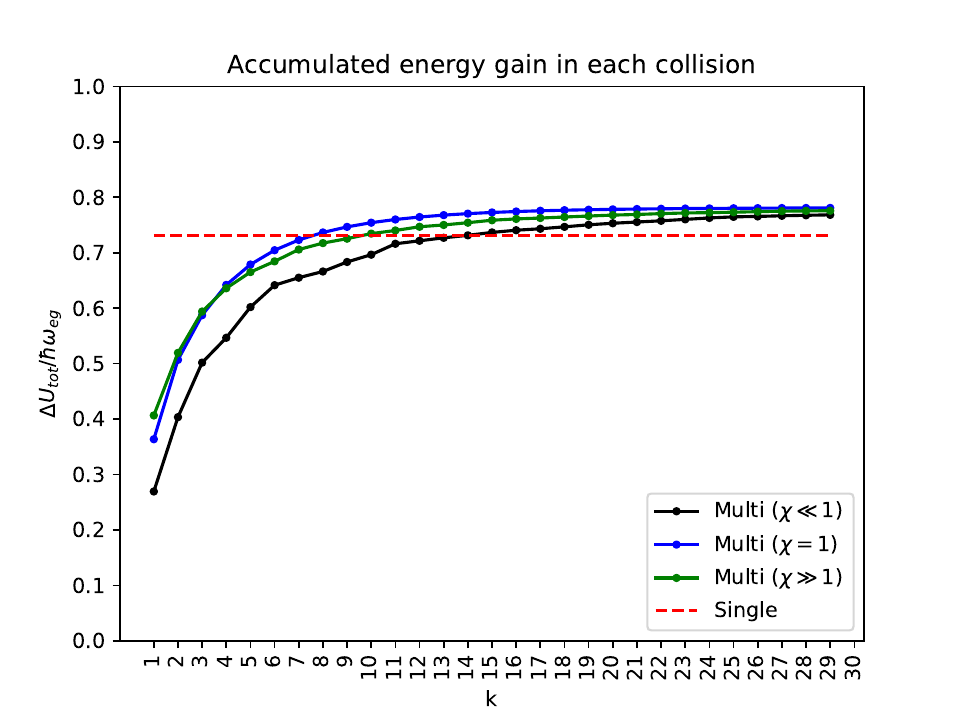}
\caption{Despite bearing the same initial mean energy, the displaced thermal state also bears coherences in its eigenenergy basis. This coherence is explored in our multi-charging protocol to achieve higher energy extraction to the string of quantum batteries.}
\label{acc_energy}
\end{figure}

One further discussion which is naturally raised in our protocol regards the main differences between the single and multi-charging scenarios. It is clear that if only one battery is considered, any NPATS ($N=1,2,...$) must always outperform the same-energy DTS whenever the selectivity tool can be applied, which can be deduced from equation (\ref{rawenergy}): displacing a thermal state only drags its vacuum population close to zero in the regime $|\alpha| \gg 1$, while for the NPATS this condition is directly met. On the other hand, coherently adding a finite number of excitations to a thermal state does not create any coherence in the free Hamiltonian eigenbasis, which happens when displacing it. We are then led to investigate whether the multi-charging protocol profits from this amount of added coherence. In figure (\ref{acc_energy}) we show the accumulated energy gain by the battery in the single-charging scheme from a SPATS (dashed, red line) \textit{versus} the accumulated energy gain by the string of batteries in the multi-charging case from a DTS (full lines with dots). Notice that despite bearing the same initial amount of internal energy, the DTS provides more charge to the string of qubits in comparison to the SPATS in the single-charging protocol. Also, notice that in the scenario where the couplings are asymmetric, the selectivity tool can be applied to the protocol and guarantee roughly the same amount of overall charging.

\begin{figure}[h!]
\centering
\includegraphics[width=\linewidth]{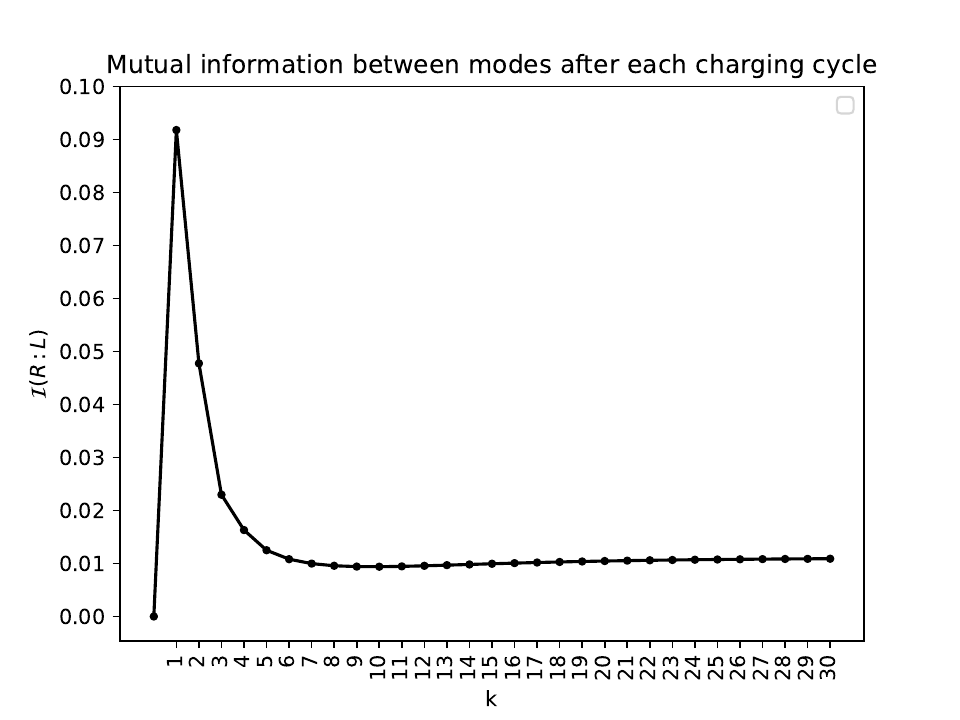}
\caption{Mutual information between both modes composing the charger in the asymptotic dynamics where $K=30$ batteries have passed through the charging cycle. The first battery, which interacts with a separable set of DTS + Gibbs states, tends to correlate them. After this, the following ones absorb part of the correlations between the modes to change their mean energy, but asymptotically some spurious correlations are no longer accessible and remain in the charger.}
\label{mutualvn}
\end{figure}

Finally, notice that the asymptotic energetic equilibration does not mean total separability in the charger's subspace. As previously stated, the condition $S_{mn} = 0$ is naturally achieved (not exclusively) if both oscillators are in a totally uncorrelated Gibbs state with same temperature. However, since each qubit interacts only with a specific partition of the bimodal Hilbert space, non-separable states $p_{m,m',n,n'}$ can still satisfy the equilibration condition $S_{mn} = 0$, and a simple way to check that is by employing the quantum mutual information $\mathcal{I}(R:L) \equiv S(\rho_{R}) + S(\rho_{L}) - S(\rho_{RL})$ between the right and left oscillators, with $S(\rho)\equiv -\Tr\rho \log\rho$ the Von Neumann entropy of system $\rho$. Indeed, as figure (\ref{mutualvn}) displays, the asymptotic behaviour of the bimodal charger is to achieve a joint state of both oscillators with still some correlations which are not used in the charging cycles.

\section{Conclusion and follow-ups}

Charging protocols aimed for two and three level quantum batteries have been proposed and studied under a variety of different circumstances. Furthermore, bimodal chargers have also been previously harnessed for increasing the charging power for qubit batteries\cite{Liwei2016, Wang2023,Alba2020,Zheng2026}.

In this work we have presented an interaction scheme between two quantum harmonic oscillators and a three-level system which, in the dispersive regime, leads to an effective three-body interaction that can be used to charge the quantum battery subspace of the qutrit. In addition, we have identified that the $\{e\leftrightarrow g,\ R\leftrightarrow L\}$ symmetry of the effective Hamiltonian also allows one to purify the qubit to its ground level with the employment of a single excitation, which can be of interest in the quantum computation area for the task of \textit{qubit resetting}. One important feature of this effective dynamics is that the vacuum population on both modes has a high impact on the performance of each task. In this paradigm, it is paramount to the protocol that the ratio between the modes' gap and the external temperature are low enough to suppress thermal excitations, which is automatically addressed in our set-up by means of high energy couplings between the systems.

We have also considered the more realistic scenario where one has imperfect preparation of the single photon added thermal state (SPATS) by benchmarking the evolution with a displaced thermal state (DTS) which is usually easier to prepare in various platforms. In addition to that, our protocol can also be applied whenever the couplings between the qutrit and both modes are asymmetric by means of a controllable classical drive which selectively picks the subspace where the energetic dynamics takes place.

We are also able to extend the protocol to furnish energy to more than one battery, which we discuss in a simple collisional model for a chain of up to $K=30$ batteries. While the single-charging scenario was (universally) optimal whenever the charger initial state contained a single excitation, the multi-charging one profits from energy coherences which are present in the DTS initial state to extract more energy to the chain when compared to the former case.

Finally, the scheme just presented can be instructive in the study of the complex quantum dynamics which arise when multiple quantum systems interact non-trivially. One direct application is on the investigation of how the presence and nature of correlations between the modes in the charger can be harnessed as resource for the charging task. In a parallel direction, it would also be interesting to characterize the asymptotic correlations left between the charger's components, and specially on possible modifications to the protocol that could further exploit them for energetic purposes.\\

\acknowledgements

This work was supported by Fundação Carlos Chagas Filho de Amparo à Pesquisa do Estado do Rio de Janeiro (FAPERJ) – project No. E-26/200.307/2023 and Conselho Nacional de Desenvolvimento Científico e Tecnológico - CNPq - Brasil - project No. 302872/2019-1.

\appendix

\section{Appendix A: General analytical results for DTS}

In this appendix we briefly derive the analytical parts of the paper, namely the expressions for the density matrix of the displaced thermal state and NPATS, as well as their purities and average excitation number.

We start by computing the matrix element of the displacement operator $\hat{D}(\alpha) = e^{-\frac{|\alpha|^{2}}{2}}e^{\alpha \hat{a}^{\dagger}} e^{-\alpha^{\ast}\hat{a}}$:

\begin{equation}
\bra{n}\hat{D}(\alpha )\ket{m} = \bra{0}\dfrac{\hat{a}^{n}}{\sqrt{n!}}e^{-\frac{|\alpha|^{2}}{2}}e^{\alpha \hat{a}^{\dagger}} e^{-\alpha^{\ast}\hat{a}}\dfrac{\hat{a}^{m\dagger}}{\sqrt{m!}}\ket{0},
\end{equation}

\noindent where we have written the states $\ket{m}$, $\ket{n}$ as the creation operators acting on the vacuum of the oscillator. In the next step, we expand the exponential $e^{-\alpha^{\ast}\hat{a}}$ in power series as

\begin{equation}
\dfrac{e^{-\frac{|\alpha|^{2}}{2}}}{\sqrt{m!n!}}\sum_{i=0}^{m}\dfrac{\left(-\alpha^{\ast}\right)^{i}}{i!}\bra{0}\hat{a}^{n}e^{\alpha\hat{a}^{\dagger}}\hat{a}^{i}\hat{a}^{m\dagger}\ket{0}.
\end{equation}

Notice that we have truncated the upper limit of summation on $i=m$, since for the rest of the range $m+1 < i < \infty$ the multiplication $\hat{a}^{i}\hat{a}^{m\dagger}\ket{0} = 0$. Performing the same step for the other exponential results in

\begin{equation}
\dfrac{e^{-\frac{|\alpha|^{2}}{2}}}{\sqrt{m!n!}}\sum_{i=0}^{m}\sum_{j=0}^{n}\dfrac{\left(-\alpha^{\ast}\right)^{i}}{i!}\dfrac{\left(\alpha\right)^{j}}{j!}\bra{0}\hat{a}^{n}\hat{a}^{j\dagger}\hat{a}^{i}\hat{a}^{m\dagger}\ket{0}.
\end{equation}

Let us now simplify the term inside the double sums. Notice that it consists of the sequence: raise by $m$, decrease by $i$, raise by $j$, and decrease by $n$. After this sequence, the $\ket{0}$ must go back to $\ket{0}$, or the inner product with $\bra{0}$ vanishes. This means that $m-i+j-n = 0$, and since $m$ and $n$ are fixed, we may use this equality to fix the only value for $j$ in the range $0 < j < n$ that satisfies such equality. Of course, this can only be done if $j = n-m+i$ is in such range; if it is not, the summation over $j$ always results in zero. This means that in order to have non-vanishing contribution to the summation we must make sure that $n-m+i \geq 0$ and $n-m+i \leq n$, which automatically tells us that $i \geq m-n$ and $i \leq m$. Since the second condition is already met in the upper limit of the summation over $i$, we must only care about the first. The final step in this analysis is to realize that if $m-n$ is negative, the range $m-n < i < 0$ is not encompassed by the summation over $i$; if it is positive, the range $0 \leq i < m-n$ does not satisfy the non-vanishing condition and therefore must be excluded from the sum over $i$. It turns out that the only non-vanishing sum over $i$ must be carried in the range $\max(0,m-n) \leq i \leq m$, to guarantee that all terms are non-null. After plugging this condition together with $j = n-m+i$ to eliminate the sum over $j$, we are left with

\begin{equation}
\begin{split}
&\dfrac{e^{-\frac{|\alpha|^{2}}{2}}}{\sqrt{m!n!}}\times\\
&\sum_{i=\max(0,m-n)}^{m}\dfrac{\left(-\alpha^{\ast}\right)^{i}}{i!}\dfrac{\left(\alpha\right)^{n-m+i}}{(n-m+i)!}\bra{0}\hat{a}^{n}\hat{a}^{(n-m+i)\dagger}\hat{a}^{i}\hat{a}^{m\dagger}\ket{0}.
\end{split}
\end{equation}

We are now able to apply the raising and lowering operators to the respective harmonic oscillator's states. We recall that

\begin{equation}
\hat{a}^{m\dagger}\ket{0} = \sqrt{m!}\ket{m},
\end{equation}

\begin{equation}
\hat{a}^{i}\ket{m} = \sqrt{\dfrac{m!}{(m-i)!}}\ket{m-i},
\label{relappend1}
\end{equation}

\begin{equation}
\hat{a}^{j\dagger}\ket{m-i} = \sqrt{\dfrac{n!}{(m-i)!}}\ket{n}\ (j=n-m+i\geq 0),
\end{equation}

\begin{equation}
\hat{a}^{n}\ket{n} = \sqrt{n!}\ket{0}.
\end{equation}

Collecting all these results and substituting in the term inside the sum, and after some algebraic reorganization:

\begin{equation}
\begin{split}
&\bra{n}\hat{D}(\alpha)\ket{m} =\\
&e^{-\frac{|\alpha|^{2}}{2}}\dfrac{\alpha^{n-m}}{\sqrt{m!n!}}\sum_{i=0}^{m}\left(-|\alpha|^{2}\right)^{i}\binom{m}{i}\binom{n}{m-i}(m-i)!.
\end{split}
\end{equation}

In this last step we have rewritten the factorials in the summation in terms of the binomial coefficient and this allows us to write the lower index of summation back to $i = 0$ since the factor $\binom{n}{m-i}$ is zero whenever $n < m-i \rightarrow i < m-n$.

With the elements $D_{nm}(\alpha)$ calculated, we are now able to write the matrix element for $\rho_{DTS} = \hat{D}(\alpha)\rho_{T}\hat{D}^{\dagger}(\alpha)$ as

\begin{equation}
\bra{n}\rho_{DTS}\ket{m} = \sum_{k=0}^{\infty}\dfrac{e^{-\beta\hbar\omega k}}{Z_0}D_{nk}(\alpha)D_{mn}(-\alpha),
\end{equation}

\noindent with $Z_0$ the partition function of the Gibbs state for a harmonic oscillator of energy gap $\hbar\omega$, and where we use the property $\hat{D}^{\dagger}(\alpha) = \hat{D}(-\alpha)$.

Notice that displacing a thermal state does not alter its purity, since the displacement operator $\hat{D}(\alpha)$ satisfies the unitary condition $\hat{D}^{\dagger}(\alpha)\hat{D}(\alpha) = \mathbb{1}$. Furthermore, the average number of photons can be computed as

\begin{equation}
\Tr\left(\hat{N}\rho_{DTS}\right) = \Tr\left(\hat{D}^{\dagger}(\alpha)\hat{N}\hat{D}(\alpha)\rho_{T}\right),
\end{equation}

\noindent if we rewrite $\hat{D}^{\dagger}(\alpha)\hat{N}\hat{D}(\alpha) = \hat{D}^{\dagger}(\alpha)\hat{a}^{\dagger}\mathbb{1}\hat{a}\hat{D}(\alpha) = \hat{D}^{\dagger}(\alpha)\hat{a}^{\dagger}\hat{D}(\alpha)\hat{D}^{\dagger}(\alpha)\hat{a}\hat{D}(\alpha)$ and use the known property $\hat{D}^{\dagger}(\alpha)\hat{a}\hat{D}(\alpha) = \hat{a} + \alpha$, which allows us to find that

\begin{equation}
\hat{D}^{\dagger}(\alpha)\hat{N}\hat{D}(\alpha) = \hat{N} + |\alpha|^{2} + \alpha\hat{a}^{\dagger} + \alpha^{\ast}\hat{a}.
\end{equation}

Since the last two terms are traceless in the multiplication for the diagonal Gibbs state, we arrive at the final result that

\begin{equation}
\langle n\rangle_{DTS} = \langle n\rangle_{T} + |\alpha|^{2},
\label{relappend3}
\end{equation}

\noindent where $\langle n\rangle_{T}$ is the average excitation number for a thermal state of temperature $T$. This is the known result that displacing a thermal state by $\alpha$ adds $|\alpha|^{2}$ excitations to the oscillator.

\section{Appendix B: General analytical results for NPATS}

The N-photon added thermal state is mathematically defined by

\begin{equation}
\rho_{N} = \dfrac{1}{Z_{N}}\hat{a}^{N\dagger}e^{-\beta\hbar\omega\hat{a}^{\dagger}\hat{a}}\hat{a}^{N},
\end{equation}

\noindent with $Z_N$ the partition function which ensures the state's normalization. This quantity can be calculated as

\begin{equation}
Z_N = \sum_{n=N}^{\infty} \bra{n}\hat{a}^{N\dagger}e^{-\beta\hbar\omega\hat{a}^{\dagger}\hat{a}}\hat{a}^{N}\ket{n} = \sum_{n=0}^{\infty} \dfrac{(n+N)!}{n!}e^{-\beta\hbar\omega n},
\end{equation}

\noindent where we have first simplified the range of summation from $n=0$ to $n=N$ because of the action of the annihilation operators on the state $\ket{n}$, used equation (\ref{relappend1}) and, in the last step, redefined the index of summation $n-N \rightarrow n$. This summation seems odd, but it may be analytically calculated by recursion. Indeed, consider the quantity $Z_{N+1}$:

\begin{equation}
\begin{split}
Z_{N+1} = &\sum_{n=0}^{\infty}\dfrac{(n+N+1)!}{n!}e^{-\beta\hbar\omega n} =\\
&\sum_{n=0}^{\infty}(n+N+1)\dfrac{(n+N)!}{n!}e^{-\beta\hbar\omega n}.
\end{split}
\end{equation}

By distributing the factors in parentheses in the sum, we find

\begin{equation}
\begin{split}
Z_{N+1} &= \sum_{n=1}^{\infty} \cancel{n}\dfrac{(n+N)!}{\cancel{n}(n-1)!}e^{-\beta\hbar\omega n} +\\
&+ (N+1)\sum_{n=0}^{\infty}\dfrac{(n+N)!}{n!}e^{-\beta\hbar\omega n}.
\end{split}
\end{equation}

Notice that in the first sum we have removed the zero associated to $n=0$. Also notice that the second sum is precisely the expression for $Z_{N}$. Now we redefine the index of summation in the first factor as $n-1 \rightarrow n$ to find

\begin{equation}
\begin{split}
Z_{N+1}&= \sum_{n=0}^{\infty} \dfrac{(n+N+1)!}{n!}e^{-\beta\hbar\omega (n+1)}\ +\ (N+1)Z_{N}\\
&= e^{-\beta\hbar\omega}Z_{N+1} + (N+1)Z_{N}.
\end{split}
\end{equation}

By noticing that $Z_0 = \left(1-e^{-\beta\hbar\omega}\right)^{-1}$, we can write the recursive relation for the partition function of the NPATS as

\begin{equation}
Z_{N} = NZ_{0}Z_{N-1},
\end{equation}

\noindent and therefore

\begin{equation}
\begin{split}
Z_{N} &= NZ_{0}\times (N-1)Z_{0}\times \cdots 2Z_{0}\times 1Z_{0}\times Z_{0}\\
&= N!Z_{0}^{N+1}.
\end{split}
\label{relappend2}
\end{equation}

Finally, we can use this expression for calculating the average number of excitations in the NPATS state:

\begin{equation}
\langle n\rangle_{N} = \Tr\left(\hat{a}^{\dagger}\hat{a}\rho_{N}\right) = \dfrac{1}{Z_{N}}\Tr\left(\hat{a}^{\dagger}\hat{a}\hat{a}^{N\dagger}e^{-\beta\hbar\omega\hat{a}^{\dagger}\hat{a}}\hat{a}^{N}\right),
\end{equation}

\noindent which using the cyclic property of the trace and the commutator between the creation and annihilation operators to write $\hat{a}^{\dagger}\hat{a} = \hat{a}\hat{a}^{\dagger} - 1$ can be recast in the form

\begin{equation}
Z_{N}\langle n\rangle_{N} = \Tr\left(\hat{a}^{N+1}\hat{a}^{N+1\dagger}e^{-\beta\hbar\omega\hat{a}^{\dagger}\hat{a}}\right) - \Tr\left(\hat{a}^{N}\hat{a}^{N\dagger}e^{-\beta\hbar\omega\hat{a}^{\dagger}\hat{a}}\right),
\end{equation}

\noindent which is similar to

\begin{equation}
\langle n\rangle_{N} = \dfrac{Z_{N+1}}{Z_{N}}\Tr\left(\rho_{N+1}\right) - \Tr\left(\rho_{N}\right) = \dfrac{Z_{N+1}}{Z_{N}} - 1.
\end{equation}

From the recursive relation (\ref{relappend2}) we finally find

\begin{equation}
\langle n\rangle_{N} = (N+1)Z_0 - 1 = NZ_{0} + \langle n\rangle_{T}.
\end{equation}

Where we use that for $N=0$ this reproduces the known result for the average occupation number of the Gibbs state $\langle n\rangle_{N=0} = \langle n\rangle_{T} = Z_0 - 1$. Notice that from this and Eq. (\ref{relappend3}) we are able to write the expression for the optimal displacement that keeps the number of excitations in the DTS similar to the NPATS:

\begin{equation}
|\alpha_{opt}|^{2} = NZ_{0}.
\end{equation}

\end{document}